\documentclass[11pt]{article}
\usepackage{amsmath}
\usepackage{amssymb}
\usepackage{latexsym}
\usepackage{color}
\usepackage{pstcol}
\usepackage{multicol}
\usepackage{graphicx}
\usepackage{amsfonts}
\newcommand{\bfi}{\bfseries\itshape}
\newcommand{\rem}[1]{}
\newcommand{\remfigure}[1]{}

\DeclareMathAlphabet{\mathbi}{OML}{cmm}{b}{it} 

\newcommand{\bel}{\begin{equation}\label}
\newcommand{\ee}{\end{equation}}
\newcommand{\ben}{\begin{enumerate}}
\newcommand{\een}{\end{enumerate}}
\newcommand{\bx}{\mbox{\boldmath$x$}}
\newcommand{\bB}{\mbox{\boldmath$B$}}
\newcommand{\bhB}{\mbox{\boldmath$\hat{B}$}}

\newcommand{\br}{\mathbi{r}}
\newcommand{\bqf}{\mathfrak{q}}
\newcommand{\bq}{\mathbi{q}}

\newcommand{\bD}{\mathbi{\mathcal{D}}}
\newcommand{\btheta}{\mbox{\boldmath$\theta$}}
\newcommand{\bu}{\mathbi{u}}
\newcommand{\bU}{\mbox{\boldmath$\mathfrak{U}$}}

\newcommand{\bv}{\mathbi{v}}
\newcommand{\bw}{\mbox{\boldmath$\omega$}}
\newcommand{\bhw}{\mbox{\boldmath$\hat{\omega}$}}
\newcommand{\bW}{\mbox{\boldmath$\mathfrak{w}$}}
\newcommand{\bWB}{\mbox{\boldmath$\mathfrak{w}$}_{B}}

\newcommand{\bchi}{\mbox{\boldmath$\chi$}}
\newcommand{\bhchi}{\boldsymbol{\hat{\chi}}}
\newcommand{\bsig}{\mbox{\boldmath$\sigma$}}

\newcommand{\bzeta}{\mbox{\boldmath$\mathfrak{q}$}}
\newcommand{\bPi}{\mbox{\boldmath$\mathfrak{P}$}}

\newcommand{\bh}{\mathbi{\hat{b}}}
\newcommand{\nh}{\mathbi{\hat{n}}}

\newcommand{\bnabla}{\mbox{\boldmath$\nabla$}}

\newcommand{\beq}{\begin{eqnarray}\label} 
\newcommand{\eeq}{\end{eqnarray}} 
\newcommand{\bc}{\begin{center}} 
\newcommand{\ec}{\end{center}} 
\newcommand{\lin}{L^{\infty}(\mathbb{D})}
\newcommand\shalf{\ensuremath{{\scriptstyle\frac{1}{2}}}}
\newcommand{\etal}{\textit{\!et~al.~}}

\newtheorem{theorem}{Theorem}

\makeatletter
\@addtoreset{equation}{section}

\makeatother

\definecolor{mush}{cmyk}{.3,.1,.6,.1}
\pagecolor{white} 
\color{black} 

\begin{document}
\bc
\textsf{\Large\bfi Quaternions and particle dynamics in the Euler fluid equations}
\ec
\bc
{\large J. D. Gibbon and D. D. Holm}
\par\vspace{2mm}
Department of Mathematics,\\Imperial College London,\\London SW7 2AZ, UK\,,
\par\vspace{3mm}
{\large R. M. Kerr}\\
\par\vspace{2mm}
Department of Mathematics,\\
University of Warwick,\\Coventry CV4 7AL, UK\,,
\par\vspace{2mm}
and\\
\par\vspace{2mm}
{\large I. Roulstone}\\
\par\vspace{2mm}
Department of Mathematics and Statistics,\\University of Surrey,\\Guildford GU2 7XH, UK\,.
\ec
\begin{abstract}\noindent
Vorticity dynamics of the three-dimensional incompressible Euler equations are cast into a 
quaternionic representation governed by the Lagrangian evolution of the tetrad consisting 
of the growth rate and rotation rate of the vorticity. In turn, the Lagrangian evolution 
of this tetrad is governed by another that depends on the pressure Hessian. Together these 
form the basis for a direction of vorticity theorem on Lagrangian trajectories. Moreover, 
in this representation, fluid particles carry ortho-normal frames whose Lagrangian evolution 
in time are shown to be  directly related to the Frenet-Serret equations for a vortex line. 
The frame dynamics suggest an elegant Lagrangian relation regarding the pressure Hessian 
tetrad. The equations for ideal MHD are similarly considered.
\end{abstract}

\section{\large Introductory and historical remarks}\label{intro}

Hamilton's determined concentration on the idea of quaternions is often depicted by mathematical 
historians as an obsession.  Lord Kelvin wrote that (O'Connor \& Robertson 1998)
\begin{quote}
\textit{Quaternions came from Hamilton after his really good work had been done, and though 
beautifully ingenious, (they) have been an unmixed evil to those who have touched them in 
any way}. 
\end{quote}
Having fallen in and out of fashion over the last century and a half (Tait 1890), quaternions 
currently play an important part in the theory of 4-manifolds, through which it has been shown 
that the essential physics of particles and fields is governed by geometric principles. Fluid 
turbulence is one of the great unsolved problems of modern science. While viscosity plays a 
dominant role in the late development of an incompressible turbulent flow through the Navier-Stokes
equations, the inviscid Euler equations determine the early and intermediate dynamics. The Euler 
fluid equations are known to be essentially geometrical, so it would not be surprising 
if quaternions were helpful in understanding their solutions.

A quaternion can be constructed from a scalar $s$ and a 3-vector $\br$ by forming the 
tetrad\footnote{We avoid the direct nomenclature ``4-vector'' because of the meaning assigned 
to this in gauge theories.} $\bqf = [s,\,\br]$ that is defined by
\bel{psm2}
\bqf = [s,\,\br] = sI - \br\cdot\bsig\,,
\ee
where $\br\cdot\bsig = \sum^{3}_{i=1} r_{i}\sigma_{i}$ and $I$ is the $2\times 2$ unit matrix. 
$\{\sigma_{1},\,\sigma_{2},\,\sigma_{3}\}$ are the Pauli spin matrices 
\bel{psm1}
\sigma_{1} = \left(\begin{array}{rr}
0 & i\\
i & 0
\end{array}\right)\,,
\hspace{2cm}
\sigma_{2} = \left(\begin{array}{rr}
0 & 1\\
-1 & 0
\end{array}\right)\,,
\hspace{2cm}
\sigma_{3} = \left(\begin{array}{rr}
i & 0\\
0 & -i
\end{array}\right)\,,
\ee
that obey the relations $\sigma_{i}\sigma_{j} = -\delta_{ij}I-\epsilon_{ijk}\sigma_{k}$\,.
A multiplication rule between two tetrads $\bqf_{1} = [s_{1},\,\br_{1}]$ and $\bqf_{2} = 
[s_{2},\,\br_{2}]$ can easily be determined from these properties
\bel{q1}
\bqf_{1}\circledast\bqf_{2} = \left[s_{1}s_{2}- \br_{1}\cdot\br_{2}\,,\, s_{1}\br_{2} + s_{2}\br_{1}
+ \br_{1}\times\br_{2}\right]\,.
\ee
This shows that quaternions are not commutative, although their associativity is easily demonstrated.
They are found to be extremely useful in modern inertial navigation systems, robotics \& graphics 
that are specifically designed to control or track rapidly moving objects undergoing three-axis 
rotations (Hanson 2006, Kuipers 1999). In fact, Hamilton discovered them in the context of an 
algorithm for rotating the telescope in his observatory. If Kelvin were alive today, he might be 
forced  to revise his negative opinion of their importance. 

Given the evidence, it is natural to reformulate Euler vorticity dynamics in terms of quaternions, 
particularly in tracking a fluid particle that carries its own ortho-normal co-ordinate system. 
Instead of setting Euler variables in standard function spaces, in which delicate geometric
information might be lost, the principal aim of this paper is to investigate the Lagrangian 
evolution of these variables in appropriate quaternionic form in order to preserve their inherent 
geometric properties. The language of quaternions thus provides us with an alternative and unique 
look at the problem of Euler vortex dynamics. These manipulations are not specifically dependent 
upon the nature of the domain $\mathbb{D}\subset\mathbb{R}^{3}$ but for those parts of our work 
where the direction of vorticity is discussed, the local existence in time of classical solutions 
is necessary (Kato 1972).  Thus we restrict $\mathbb{D}$ to a three-dimensional periodic domain, 
although other more general forms of $\mathbb{D}\subset\mathbb{R}^{3}$ are also valid (see Majda 
\& Bertozzi 2001). Otherwise our manipulations should be considered to be formal, particularly 
since Euler data gets rough very quickly.

Three-dimensional Euler vorticity growth is driven by the stretching vector $\bw\cdot\nabla\bu$. 
This term plays a fundamental role in determining whether or not a singularity forms in finite 
time. Major computational 
studies can be found in Brachet \etal (1983,\,1992); Pumir \& Siggia (1990); Kerr (1993,\,2005); 
Grauer \etal (1998), Pelz (2001) and Hou \& Li (2006). The Beale-Kato-Majda theorem (Beale \etal 
1982) has been the main cornerstone of Euler analysis\,: one version of this theorem is the precise
statement that $\int_{0}^{t}\|\bw\|_{\lin}d\tau$ must be finite to prevent singular behaviour on 
$\mathbb{D}$. A BMO-version of this theorem has been proved by Kozono and Taniuchi (2000). However,
it has become clear that not only the magnitude but also the direction 
of vorticity is important. The papers by Constantin (1994), Constantin \etal (1996), Cordoba \& 
Fefferman (2001), Deng \etal (2004,\,2005) and Chae (2003,\,2005,\,2006) are variations on this 
theme. References and a more global perspective on the Euler equations can be found in the book 
by Majda and Bertozzi (2001). Shnirelman (1997) has constructed very weak solutions which have 
some realistic features but whose kinetic energy monotonically decreases in time and which are 
everywhere discontinuous and unbounded. For work on Euler limits see Brenier (1999,\,2000) and 
for its dynamics in the more exotic function spaces see the papers by Tadmor (2001) and Chae 
(2003,\,2004). 

The new results in this paper displayed in Sections 1-6 can be summarized as follows.

A well-known variable is the scalar growth rate $\alpha = \bhw\cdot S\bhw$ (Constantin 1994).
Associated with this is the 3-vector rotation or swing rate $\bchi = \bhw\times S\bhw$\,, where 
$\bhw$ is 
the unit vorticity and $S = \shalf(u_{i,j}+u_{j,i})$ is the strain matrix. Together these form a 
natural tetrad\footnote{In Gibbon (2002) $\bzeta = [\alpha,\,\bchi]$ was denoted as 
$\mbox{\boldmath$\zeta$}$\,. The change of notation to Gothic variables for tetrads has been 
introduced to avoid confusion between these and 3-vectors.} $\bzeta = [\alpha,\,\bchi]$. 
Theorem 1 of \S2 shows that the Lagrangian advection equation for the vorticity tetrad 
$\bW =[0,\,\bw]$ can then be written as
\bel{wequn}
\frac{D\bW}{Dt} = \bzeta\circledast\bW\,.
\ee 
All these quaternionic variables are Eulerian variables\,; i.e., point-wise functions of space 
and time, but undergoing Lagrangian evolution in time. 

The tetrad $\bzeta$ satisfies its own Lagrangian advection equation driven by the effect of 
the pressure Hessian $P = \{p_{,ij}\}$ through the variables $\alpha_{p} = \bhw\cdot P\bhw$ and 
$\bchi_{p} = \bhw\times P\bhw$. Together these also form a natural
tetrad  $\bzeta_{p} = [\alpha_{p},\,\bchi_{p}]$. Figure 1 shows how $S\bw$, $P\bw$ and the 
three orthonormal vectors $(\bhw,\,\bhchi,\,\bhw\times\bhchi)$ are related. In addition to 
(\ref{wequn}), Theorem 1 also contains the results for the Lagrangian advection of $\bzeta$ 
and $\bzeta_{p}$.  Simply stated this is
\bel{zintro1a}
\frac{D\bzeta}{Dt} + \bzeta\circledast\bzeta + \bzeta_{p} = 0\,.
\ee

The result in (\ref{zintro1a}) enables us to prove a Theorem in \S\ref{vort} on the direction of 
vorticity based on Lagrangian trajectories\,: ``\textit{Provided $\|\bchi_{p}\|_{\lin}$ is 
integrable in time up to $t^* > 0$ on a periodic domain $\mathbb{D}$, no Euler singularity is 
possible at $t^*$, with the exception of the case where $\bhw$ becomes collinear with an 
eigenvector of $P$ at $t^*$}''. Although different in detail, this 
result is in the same style as the direction of vorticity theorems cited above and is directly 
a variation of the BKM theorem. Ohkitani and Kishiba (1995) have observed in computations 
that at maximum points of enstrophy, $\bw$ becomes collinear with the most negative eigenvector 
of $P$. Collinearity may therefore be an important process in vorticity growth. The pressure 
Hessian $P$ and its interplay with the strain matrix $S$ has appeared in the Euler and Navier-Stokes
literature in various places; see the references in Galanti \etal (1997), Majda and Bertozzi 
(2001) and Chae (2006).

At each point in space-time a fluid particle carries its own ortho-normal co-ordinate 
system $(\bhw,\,\bhchi,\,\bhw\times\bhchi)$\,: see Figure 1. Explicit equations for 
Lagrangian time derivatives of this frame are given in \S3. The corresponding Darboux 
vector is the particle rotation rate. The frame-equations are then shown to be 
directly related to the Frenet-Serret relations of differential geometry that govern 
the curvature and torsion of a vortex line through the arc-length derivative of its 
tangent, principal unit normal and bi-normal. Using Ertel's Theorem, explicit 
differential equations for the curvature and torsion are then found.

It is shown in \S\ref{conjec} how to find Lagrangian differential equations for 
$\alpha_{p}$ and $\bchi_{p}$.  The relation between $\bzeta_{p}$ and $\bzeta$ is given 
in Theorem \ref{zetapthm} where they are shown to satisfy 
\bel{zintro1b}
\frac{D\bzeta_{p}}{Dt} = \bzeta\circledast\bzeta_{p} + \bPi\,.
\ee
$\bPi$ is a tetrad linear in $\bzeta$ and $\bzeta_{p}$ whose arbitrary scalar coefficients, in 
principle, are determined by the Poisson pressure relation. 

The vorticity vector-field $\bw\cdot\nabla$ is frozen into the Euler flow. Any system 
with a frozen-in vector-field will also have an associated form of Ertel's Theorem, and a 
corresponding tetrad $\bzeta = [\alpha,\,\bchi]$. Thus the Lagrangian-quaternionic format 
displayed in this paper is more generally applicable, as illustrated by the equations for 
ideal MHD in \S \ref{MHD}. Two time-clocks and two tetrads $\bzeta^{\pm} = 
[\alpha^{\pm},\,\bchi^{\pm}]$ appear as a result because of the two Lagrangian 
derivatives that naturally arise through the use of Elsasser variables. 

Previous attempts at formulating Euler vorticity dynamics using quaternions have met with 
only partial success. Past results have appeared in reverse order\,: the relations between 
$\alpha$ and $\bchi$ to be displayed in Theorem 1 were derived first by Galanti \etal (1997) 
(see also Gibbon \etal 2000), which were then shown to be expressible in a quaternionic form 
(Gibbon 2002). That 
story was incomplete, however, because the Lagrangian advection equation for $\bW$ was 
missing, as were the ideas on particle frame dynamics, the pressure relation (\ref{zintro1b}), 
and results on the direction of vorticity. Roubtsov \& Roulstone (1997,\,2001) 
have also formulated semi-geostrophic theory in terms of quaternions. 

\section{\large Vorticity dynamics in quaternion form}\label{vort}

\par\vspace{2mm}\noindent
\bc
\begin{minipage}[c]{.45\textwidth}
\begin{pspicture}
\psframe(0,0)(5,5)
\thicklines
\qbezier(0,0)(4,2.5)(0,5)
\thinlines
\put(2,2.5){\vector(0,1){1}}
\put(2,3.8){\makebox(0,0)[b]{$\bhw$}}
\thicklines
\put(2,2.5){\vector(-1,1){1}}
\put(.5,3.3){\makebox(0,0)[b]{$S\bw$}}
\thinlines
\put(2,2.5){\vector(3,1){1}}
\put(3.2,3.0){\makebox(0,0)[b]{$\bhw\times\bhchi$}}
\put(2,2.5){\vector(1,0){1}}
\put(3.8,2.4){\makebox(0,0)[b]{$\bhchi$}}
\end{pspicture}
\end{minipage}
\begin{minipage}[c]{.45\textwidth}
\begin{pspicture}
\psframe(0,0)(5,5)
\thicklines
\qbezier(0,0)(4,2.5)(0,5)
\thinlines
\put(2,2.5){\vector(0,1){1}}
\put(2,3.8){\makebox(0,0)[b]{$\bhw$}}
\thicklines
\put(2,2.5){\vector(-2,3){0.8}}
\put(.9,3.8){\makebox(0,0)[b]{$P\bw$}}
\thinlines
\put(2,2.5){\vector(3,-2){1}}
\put(3.3,1.6){\makebox(0,0)[b]{$\bhchi_{p}$}}
\put(2,2.5){\vector(4,1){1}}
\put(3.8,2.6){\makebox(0,0)[b]{$\bhw\times\bhchi_{p}$}}
\end{pspicture}
\end{minipage}
\ec
\bc
\begin{minipage}[r]{\textwidth}
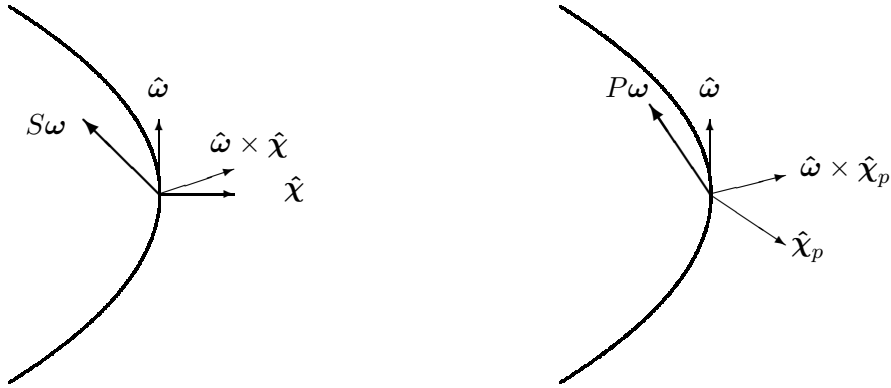
{\small \textbf{Figure 1:} A vortex line with unit tangent vorticity vector $\bhw$. 
The normal vectors $\bchi = \bhw\times S\bhw$ and $\bchi_{p} = \bhw\times P\bhw$ 
are defined in (\ref{alphachidef1}) \& (\ref{alphachidef2}). Thus the three unit 
vectors $[\bhw,\,\bhchi,\,\bhw\times\bhchi]$ form an ortho-normal co-ordinate system. 
Moreover, $\bhw$, $S\bw$ and $\bhw\times\bhchi$ are co-planar, as are $\bhw$, $P\bw$ 
and $\bhw\times\bhchi_{p}$.}
\end{minipage}
\ec
\par\vspace{0mm}\noindent
In their vorticity form, the three-dimensional incompressible Euler equations are
\bel{eul1a}
\frac{D\bw}{Dt} = \bw\cdot\nabla\bu = S\bw\,,
\ee
where the strain matrix is written as $S = \shalf\left(u_{i,j}+u_{j,i}\right)$ and $\bw 
= \mbox{curl}\,\bu$ is the vorticity (Majda and Bertozzi 2001). Equation (\ref{eul1a}) 
arises from taking the curl of the Euler equations in their velocity formulation
\bel{eul1b}
\frac{D\bu}{Dt} = -\nabla p\,,\hspace{3cm}\rm{div}\,\bu = 0\,,
\ee
in which the Lagrangian (material) derivative is defined as
\bel{eul1c}
\frac{D~}{Dt} = \frac{\partial~}{\partial t} + \bu\cdot\nabla\;.
\ee
The vorticity can be expressed as a tetrad by taking the quaternionic curl of $\bU = [0,\,\bu]$
\bel{wtetdef1}
\bnabla\circledast\bU = [-\hbox{div}\,\bu,\, \hbox{curl}\,\bu]\,.
\ee
Thus there exists a natural vorticity tetrad $\bW$ has the divergence-free constraint built 
into it
\bel{wtetdef2}
\bW = [0,\,\bw]\,.
\ee
The results in this paper employ Ertel's theorem, which is widely used in geophysical 
fluid dynamics in the study of potential vorticity\,: see Hide (1983, 2004) and Hoskins, 
\etal (1985). More generally it applies to any fluid system whose flow preserves a vector 
field, as the Euler equations preserve $\bw\cdot\nabla$. For the extensive history behind 
this result, which seems to have originated with Cauchy, see Truesdell \& Toupin (1960), 
Kuznetsov and Zakharov (1997) and Viudez (1999). The most general form of Ertel's Theorem 
says that if $\bw$ satisfies (\ref{eul1a}) then for an arbitrary differentiable vector 
$\btheta$
\bel{ertel1}
\frac{D~}{Dt}\left(\bw\cdot\nabla\btheta\right) = \bw\cdot\nabla\left(
\frac{D\btheta}{Dt}\right)\,.
\ee
The choice of $\btheta$ as the Euler velocity field $\bu$ (Ohkitani 1993) implies that 
the vortex stretching vector $\bw\cdot\nabla\bu = S\bw$ is governed by 
\bel{ertel2}
\frac{D(S\bw)}{Dt} = - P\bw\,,
\ee
where $P = \{p_{,ij}\}= \left\{\partial^{2} p/\partial x_{i}\partial x_{j}\right\}$ is the 
Hessian matrix of the pressure. Thus the combination of  (\ref{eul1a}) and (\ref{ertel2}) 
gives Ohkitani's relation (Ohkitani 1993)
\bel{ertel3}
\frac{D^{2}\bw}{Dt^{2}} + P\bw = 0\,.
\ee
To understand how the \textit{direction} in which the vorticity vector stretches (compresses) in 
relation to its growth rate requires an understanding of its relationship with the matrices $S$ 
and $P$.  The scalar and vector variables $\alpha$ and $\bchi$ are defined by
\bel{alphachidef1}
\alpha = \bhw\cdot S\bhw\,,\hspace{3cm}\bchi = \bhw\times S\bhw\,,
\ee
\bel{alphachidef2}
\alpha_{p} = \bhw\cdot P\bhw\,,\hspace{3cm}\bchi_{p} = \bhw\times P\bhw\,.
\ee
The left part of Figure 1, based upon $S\bhw$, shows the ortho-normal co-ordinate system 
$\bhw$, $\bhchi$ and $\bhw\times\bhchi$; the right hand part of the figure shows the same 
figure with $S$ replaced by $P$. Thus $S\bhw$ can be decomposed into its components along 
the two orthogonal vectors $\bhw$ and $\bchi\times\bhw$
\bel{eul3}
S\bhw  = \alpha\,\bhw + \bchi\times \bhw\,.
\ee
From (\ref{eul1a}) and (\ref{eul3}), the Lagrangian derivatives of $|\bw|$ 
and $\bhw$ are given by
\bel{eul4a}
\frac{D|\bw|}{Dt} = \alpha\,|\bw|\,,
\hspace{3cm}
\frac{D\bhw}{Dt} = \bchi\times\bhw\,.
\ee
The quantities $(\alpha,\,\bchi)$ are respectively the rates of change in vorticity 
magnitude and direction; that is, one may call  respectively call $\alpha$ and $\bchi$ 
the stretching rate\footnote{$\alpha$ and $\alpha_{p}$ are Rayleigh quotient estimates 
for eigenvalues of $S$ and $P$ respectively although they are only exact eigenvalues 
when $\bw$ aligns with one of their eigenvectors. Constantin (1994) has a Biot-Savart 
formula for $\alpha$.} and the rotation or swing rate. These variables form natural 
tetrads associated with $\bW = [0,\,\bw]$
\bel{z1}
\bzeta = \left[\alpha\,,\bchi\right]\,,
\hspace{2cm}
\bzeta_{p} = \left[\alpha_{p}\,,\bchi_{p}\right]\,.
\ee
The following theorem shows how Euler vorticity dynamics can be formulated using quaternions.
\par\medskip\noindent
\begin{theorem}\label{thm1}{\bf[Euler vorticity dynamics in terms of quaternions:]}
The vorticity tetrad $\bW(\bx,t)$ satisfies the relation
\bel{l1a}
\frac{D\bW}{Dt} = \bzeta\circledast\bW\,,
\ee
while Ohkitani's relation (\ref{ertel3}) becomes
\bel{l1b}
\frac{D^{2}\bW}{Dt^{2}} + \bzeta_{p}\circledast\bW = 0\,.
\ee
The tetrad $\bzeta(\bx,t)$ satisfies the compatibility relation (Riccati equation)
\bel{l2}
\frac{D\bzeta}{Dt} + \bzeta\circledast\bzeta + \bzeta_{p} = 0\,.
\ee
\end{theorem}
\par\vspace{2mm}\noindent
\textbf{Remark:} In terms of $\alpha$ and $\bchi$, the components of (\ref{l2}) were originally 
calculated by an indirect route in Gibbon (2002), although at that time (\ref{l1a}) was not yet 
available. Moreover the present formulation simplifies the proof.
\par\vspace{2mm}\noindent
\textbf{Proof:} (\ref{l1a}) follows from (\ref{eul1a}) and (\ref{eul3}) by direct calculation
\bel{p1}
\frac{D\bW}{Dt} = \left[0,\,\alpha \bw + \bchi\times\bw\right] = \left[\alpha\,,\bchi\right]
\circledast
\left[0,\,\bw\right] = \bzeta\circledast\bW\,.
\ee
Following (\ref{eul3}) and Figure 1, we have 
\bel{p2a}
P\bw = \alpha_{p}\bw + \bchi_{p}\times\bw
\hspace{1cm}\Rightarrow\hspace{1cm}
[0,\,P\bw] = \bzeta_{p}\circledast\bW\,.
\ee
Consequently, Ohkitani's relation (\ref{ertel3}) implies
\bel{p2b}
\frac{D^{2}\bW}{Dt^{2}}  = \frac{D~}{Dt}[0,\,S\bw] = -[0,\,P\bw]
= - \bzeta_{p}\circledast\bW\,,
\ee
which is (\ref{l1b}). Differentiating (\ref{l1a}) again and using (\ref{l1b}) gives the 
compatibility relation
\bel{p3}
\frac{D\bzeta}{Dt}\circledast\bW + \bzeta\circledast(\bzeta\circledast\bW)+ 
\bzeta_{p}\circledast\bW = 0\,.
\ee
The result (\ref{l2}) in Theorem \ref{thm1} follows because of the associativity property.
\hspace{2cm}$\blacksquare$
\par\medskip
The meaning of $\bchi$ now becomes clear. For structures such as straight vortex tubes or flat 
sheets, $\bw$ aligns with an eigenvector of $S$ and thus $\bchi = 0$, in which case $\alpha$ is 
an exact eigenvalue of $S$.  The Ricatti equation for $\bzeta$ in (\ref{l2}) reduces to a simple 
scalar form. However as soon as a tube or sheet bends, twists or tangles, $\bchi \neq 0$ and 
the full tetrad is restored. Because all our variables are functions of $(\bx\,,t)$, equations 
(\ref{l1a}) and (\ref{l2}) govern the vorticity dynamics at all points and all times in the flow 
provided solutions remain finite.
\par\smallskip
The BKM-theorem (Beale \etal 1984) shows that the time integral 
$\int_{0}^{t^*}\|\bw\|_{\lin}\,d\tau$ must be finite at a time $t^*$ to rule out singular 
behaviour. Variations on this theme are the direction of vorticity theorems expressed in the 
work of Constantin \etal (1996),  Cordoba \& Fefferman (2001), Deng \etal (2004,\,2005) and 
Chae (2006). Chae's result (his Theorem 5.1) is based on control of the time integral of 
$\|S\bhw\cdot P\bhw\|_{\infty}$, which is derivable from (\ref{ertel2}).  Here, a direct 
consequence of Theorem \ref{thm1} concerns the pressure Hessian and its associated variable 
$\bchi_{p}$\,.
\par\smallskip\noindent
\begin{theorem}\label{chipthm}{\bf\,:} On the domain $\mathbb{D} = [0,L]^{3}_{per}$ there 
exists a global solution of the Euler equations, $\bu \in C([0,\,\infty];H^{s})
\cap C^{1}([0,\,\infty];H^{s-1})$ for $s\geq 3$ if, for every $t^{*}>0$
\bel{integ1b}
\int_{0}^{t^*}\|\bchi_{p}\|_{\lin}\,d\tau < \infty\,,
\ee
excepting the case where $\bhw$ becomes collinear with an eigenvector of $P$ at $t^*$.
\end{theorem}
\par\small\noindent
\textbf{Remark:} The theorem does not imply that blow-up occurs when collinearity does; it
simply implies that under condition (\ref{integ1b}) it is the only situation when it can happen.
Ohkitani (1993) and Ohkitani and Kishiba (1995) have noted the collinearity mentioned above; 
they observed in Euler computations that at maximum points of enstrophy, $\bw$ tends to align 
with the eigenvector corresponding to the most negative eigenvalue of $P$. 
\par\medskip\noindent
\textbf{Proof:} Consider the relation for $\bzeta$ in Theorem \ref{thm1} in (\ref{l2}) expressed 
in $\alpha$--$\chi$ components 
\bel{ac2}
\frac{D\alpha}{Dt} = \chi^2 - \alpha^2 -\alpha_{p}
\hspace{2cm}
\frac{D\chi}{Dt} = -2\alpha\chi - \bhchi\cdot\bchi_{p}
\ee
where $\alpha = \alpha(x,t)$ and $\bchi(t)= \bchi(x,\,t)$. From (\ref{eul3}) we know that 
\bel{ac3}
|S\bhw|^2 = \alpha^2 + \bchi^2
\ee
and so
\bel{ac3}
\frac{D|S\bhw|}{Dt}\leq -\alpha |S\bhw| + \frac{|\alpha||\alpha_{p}| 
+ |\bchi||\bchi_{p}|}{(\alpha^2 + \bchi^2)^{1/2}}\,.
\ee
Because $D|\bw|/Dt = \alpha|\bw|$ from (\ref{eul4a}), the magnitude of vorticity $|\bw|$ cannot 
blow-up for $\alpha < 0$. Thus our concern is with $\alpha \geq 0$. Inequality (\ref{ac3}) 
becomes
\bel{ac4}
\frac{D|S\bhw|}{Dt}\leq |\alpha_{p}| + |\bchi_{p}|\,.
\ee
Integrating along bounded Lagrangian trajectories $X(t,{\small\bx_{0})}$ that satisfy $X_{t} = 
u(X(t,{\small\bx_{0})},t)$, we have 
\bel{ac5}
S\bw(X(t,{\small\bx_{0})},t) = S\bw(X(0,{\small\bx_{0})},0) + 
\int_{0}^{t}\left(  |\alpha_{p}(X(\tau,{\small\bx_{0})},\tau)| 
+ |\bchi_{p}(X(\tau,{\small\bx_{0})},\tau)|\right)\,d\tau\,.
\ee
When there is no collinearity between $\bhw$ and $P\bhw$, the assumption of point-wise in space 
integrability in time of $\bchi_{p}$ in (\ref{integ1b}) also extends to $\alpha_{p}$. Thus 
$\|S\bhw\|_{\infty}$ is bounded if (\ref{integ1b}) holds and, in consequence, so is $|\bw|$. 
The Beale-Kato-Majda theorem then guarantees regularity of the Euler equations. 
However, there still exists the possibility that 
$|P\bhw|$ could blow up simultaneously as the 
angle between $\bhw$ and $P\bhw$ approaches zero while keeping $\bchi_{p}$ finite; under these 
circumstances $\int_{0}^{t}\|\bchi_{p}\|_{\lin}d\tau < \infty$, whereas 
$\int_{0}^{t}\|\alpha_{p}\|_{\lin}d\tau \to\infty$; thus blow-up would still be theoretically 
possible. \hspace{3cm}$\blacksquare$


\section{\large Lagrangian frame dynamics for particles and the Frenet-Serret equations}\label{frame}

\subsection{Frame dynamics for particles}

The Lagrangian dynamics of the ortho-normal frame $(\bhw,~\bhchi,~\bhw\times\bhchi)$ can now 
be evaluated. Figure 2 illustrates the motion of a particle from one co-ordinate point in 
space-time to another
\par\vspace{.25cm}\noindent
\bc
\begin{minipage}[c]{.75\textwidth}
\begin{pspicture}
\psframe(0,0)(5,5)
\thicklines
\qbezier(0,0)(4,2.5)(0,5)
\thinlines
\put(1,.1){\makebox(0,0)[b]{$t_{1}$}}
\put(2.01,2.43){\makebox(0,0)[b]{$\bullet$}}
\put(1.4,2.5){\makebox(0,0)[b]{\scriptsize$(\bx_{1},t_{1})$}}
\thinlines
\put(2,2.5){\vector(0,1){1}}
\put(2,3.7){\makebox(0,0)[b]{$\bhw$}}
\put(2,2.5){\vector(-2,-1){1}}
\put(.7,1.8){\makebox(0,0)[b]{$\bhchi$}}
\put(2,2.5){\vector(1,0){1}}
\put(3.8,2.4){\makebox(0,0)[b]{$\bhw\times\bhchi$}}
\thicklines
\qbezier(7,0)(6,2.5)(8,5)
\thinlines
\put(7.3,.1){\makebox(0,0)[b]{$t_{2}$}}
\put(6.77,2.45){\makebox(0,0)[b]{$\bullet$}}
\put(6.2,2.5){\makebox(0,0)[b]{\scriptsize$(\bx_{2},t_{2})$}}
\thicklines
\qbezier[50](2,2.5)(3,.5)(6.7,2.5)
\thinlines
\put(6.73,2.5){\vector(1,4){.3}}
\put(7,3.8){\makebox(0,0)[b]{$\bhw$}}
\put(6.7,2.5){\vector(4,-1){1}}
\put(8.4,2.8){\makebox(0,0)[b]{$\bhw\times\bhchi$}}
\put(6.7,2.5){\vector(4,1){1}}
\put(8.1,2.1){\makebox(0,0)[b]{$\bhchi$}}
\put(3,1.2){\vector(1,0){.6}}
\put(3.5,.7){\makebox(0,0)[b]{\small Particle trajectory}}
\put(4.5,1.4){\vector(4,1){.6}}
\thinlines
\end{pspicture}
\end{minipage}
\ec
\bc
\vspace{.25cm}
\begin{minipage}[r]{\textwidth}
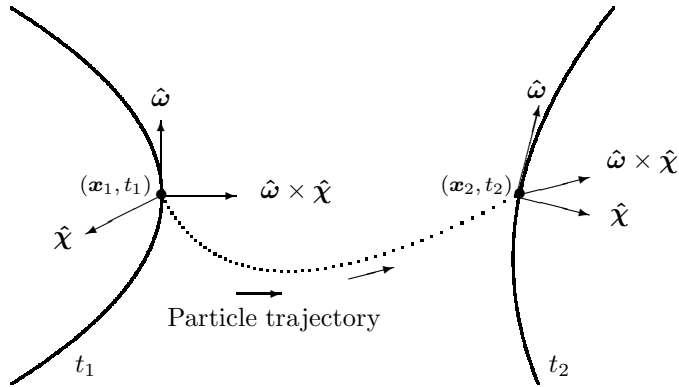
\textbf{Figure 2:} {\small Vortex lines at two different times $t_{1}$ and $t_{2}$, with the dotted
line representing the particle  $(\bullet)$ trajectory moving from $(\bx_{1},t_{1})$ to $(\bx_{2},t_{2})$.
The orientation of the ortho-normal unit vectors $(\bhw,\bhchi,~\bhw\times\bhchi)$ is shown at each
of the two space-time points.}
\end{minipage}
\ec
To find a closed expression for the Lagrangian time derivatives of the ortho-normal set 
$(\bhw,\,\bhchi,\,\bhw\times\bhchi)$ requires the derivative of $\bhchi$. To find 
this it is necessary to use the fact that the 3-vector $P\bhw$ can be expressed in this 
ortho-normal frame as the linear combination
\bel{orthog1}
P\bhw = \alpha_{p}\,\bhw + c_{1}\bhchi + c_{2}(\bhw\times\bhchi)\,.
\ee
where the coefficients $c_{1}$ and $c_{2}$ are defined by 
\bel{c12defa}
c_{1} = \bhw\cdot(\bhchi\times\bchi_{p})
\hspace{2.5cm}
c_{2} = -\, (\bhchi\cdot\bchi_{p})
\ee
The 3-vector product $\bhw\times P\bhw$ yields 
\bel{orthog2}
\bchi_{p} = c_{1}(\bhw\times\bhchi) - c_{2}\bhchi\,.
\ee
The Lagrangian time derivative of $\bhchi$ comes from the 3-vector part of equation (\ref{l2}) 
for the tetrad $\bzeta = [\alpha,\,\bchi]$ in Theorem \ref{thm1}
\bel{frame1a}
\frac{D\bchi}{Dt} = - 2\alpha\bchi - \bchi_{p}
\hspace{1cm}
\Rightarrow
\hspace{1cm}
\frac{D\chi}{Dt} = -2\alpha \chi + c_{2}\,,
\ee
where $\chi = |\bchi|$. Using (\ref{orthog2}) and (\ref{frame1a}) there follows
\bel{frame1b}
\frac{D\bhchi}{Dt} = - c_{1}\chi^{-1}(\bhw\times\bhchi)\,,
\hspace{2cm}
\frac{D(\bhw\times\bhchi)}{Dt} = \chi\,\bhw + c_{1}\chi^{-1}\bhchi\,.
\ee
Thus, according to Euler's fluid equations, the Lagrangian time derivatives of the 
ortho-normal set can be expressed as
\beq{frameA1}
\frac{D\bhw}{Dt}&=& \bD\times\bhw\\
\frac{D(\bhw\times\bhchi)}{Dt} &=& \bD\times(\bhw\times\bhchi)\label{frameA2}
\\
\frac{D\bhchi}{Dt} &=& \bD\times\bhchi\label{frameA3}
\eeq
where the ``Darboux angular velocity vector'' $\bD$ for the ortho-normal frame is defined as
\bel{frame2}
\bD = \bchi - \frac{c_{1}}{\chi}\boldsymbol{\hat{\omega}}
\hspace{1.5cm}\hbox{with}\hspace{1.5cm}
|\bD|^2 = \chi^{2} + \frac{c_1^2}{\chi^2}\,.
\ee

\subsection{\large Frame dynamics and the Frenet-Serret equations}\label{vortex-line-dyn}

\par\vspace{2mm}\noindent
\bc
\begin{minipage}[c]{.45\textwidth}
\begin{pspicture}
\psframe(0,0)(5,5)
\thicklines
\qbezier(0,0)(4,2.5)(0,5)
\thinlines
\put(2,2.5){\vector(0,1){1}}
\put(2,3.8){\makebox(0,0)[b]{$\bhw$}}
\thicklines
\thinlines
\put(2,2.45){\vector(1,0){1}}
\put(4,2.4){\makebox(0,0)[b]{$\bhw\times\bhchi = \nh$}}
\put(2,2.45){\vector(3,2){1}}
\put(3.6,3.1){\makebox(0,0)[b]{$\bhchi = \bh$}}
\end{pspicture}
\end{minipage}
\par\vspace{5mm}\noindent
\begin{minipage}[r]{\textwidth}
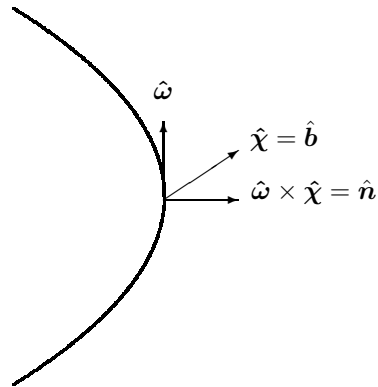
\textbf{Figure 3:} {\small The ortho-normal frame $(\bhw,~\bhw\times\bhchi,~\bhchi)$
as the Frenet-Serret frame.}
\end{minipage}
\ec
With $\bhw$ as the unit tangent vector, $\bhchi$ as the unit bi-normal $\bh$ and 
$\bhw\times\bhchi$ as the unit principal normal $\nh$, the matrix $F$ can be formed
\bel{frmx}
F = \left(\bhw^{T},\,(\bhw\times\bhchi)^{T},\,\bhchi^{T}\right)\,,
\ee
and (\ref{frameA1})--(\ref{frameA3}) can be re-written as
\bel{frame3a}
\frac{DF}{Dt} = AF\,,\hspace{2cm}
A = \left(
\begin{array}{ccc}
0 &-\chi & 0\\
\chi & 0 &-c_{1}\chi^{-1}\\
0& c_{1}\chi^{-1} & 0
\end{array}\right)\,.
\ee
For a space curve parameterized by arc-length $s$, 
then the Frenet-Serret equations relating $dF/ds$ to the curvature $\kappa$ and the torsion $\tau$ 
of a vortex line are
\bel{frame3b}
\frac{dF}{ds} = BF
\hspace{1.5cm}\hbox{where}
\hspace{1.5cm}
B = \left(
\begin{array}{rrr}
0    & \kappa     & 0\\
-\kappa & 0         & \tau\\
0    & -\tau & 0
\end{array}
\right)\,.
\ee
It is now possible to relate the $t$ and $s$ derivatives of $F$ given in (\ref{frame3a}) and 
(\ref{frame3b}). At any time $t$ the integral curves of the vorticity vector field define a 
space-curve through each point $\bx$: these space curves are called `vortex lines'. The 
arc-length derivative $d/ds$ is defined by
\bel{integ1a}
\frac{d}{ds} = \bhw\cdot\nabla\,.
\ee
The evolution of the curvature $\kappa$
and torsion $\tau$ of a vortex line may be obtained from Ertel's theorem in (\ref{ertel1}), 
expressed as the commutation of operators
\bel{frame3d}
\alpha \frac{d~}{ds} + \Big[\frac{D}{Dt},\,\frac{d}{ds}\Big] = 0\,.
\ee
Applying this to $F$ and using the relations (\ref{frame3a}) and (\ref{frame3b}) gives the 
Lax pair
\bel{com1}
\alpha B+ \frac{DB}{Dt} - \frac{dA}{ds} + [B,\,A] = 0\,.
\ee
Thus Ertel's Theorem gives explicit evolution equations for the curvature $\kappa$ and 
torsion $\tau$ that lie within the matrix $B$ and relates them to $c_{1}$ and $\chi$.

Finally we remark that the frame dynamics along each Lagrangian trajectory may be characterized 
by a curve in the $\mathbb{R}^{2}$ or $\mathbb{C}^{1}$ plane. For example, one might consider 
the quantity 
\bel{has1}
\psi (s,t) = |\mathcal{D}| \exp\left(i\int^{t} [c_{1}\chi^{-1}](s,t')\,dt'\right) 
\ee
evaluated along each Lagrangian trajectory. This complex representation of the Darboux
vector's effect is reminiscent of the Hasimoto transformation 
\bel{has2}
\psi (s,t) = \kappa(s,t) \exp\left(i\int^{s} \tau(s',t)\,ds'\right) 
\ee
used for representing the 
propagation of a Kelvin wave along a vortex filament in terms of its induced curvature 
and torsion (Hasimoto 1972). Such a representation is potentially useful as a diagnostic 
for characterizing frame dynamics in an experimental or computational fluid flow. Thus, 
because the Darboux vector has only two components, a representation exists that reduces 
the description of frame rotation for each fluid element to a curve in a plane. 

\section{\large A Lagrangian advection equation for $\bzeta_{p}$}\label{conjec}

One of the hurdles in pursuing a Lagrangian approach to the Euler equations is the problem of 
the non-locality of the pressure field. Overtly, we have no Lagrangian differential equations 
for either $\alpha_{p}$ or $\bchi_{p}$: the usual numerical procedure is to up-date the pressure 
through its Poisson equation $-\Delta p = u_{i,j}u_{j,i}$. How to address this issue can be 
illustrated by an example. Differentiating the orthogonality relation $\bchi\cdot\bhw = 0$ 
and using the derivative of $\bhw$ in (\ref{eul4a}) gives
\bel{press1}
\bhw\cdot\frac{D\bchi}{Dt} = 0
\hspace{1cm}
\Rightarrow
\hspace{1cm}
\frac{D\bchi}{Dt} = \bq_{0}
\ee
where $\bq_{0}$ lies in the plane perpendicular to $\bhw$ in which $\bchi$ and $\bchi_{p}$ 
also lie. Thus $\bq_{0} = A(\bx,t)\,\bchi + B(\bx,t)\,\bchi_{p}$\,; independently it is known 
that $A=-2\alpha$ and $B=-1$ from (\ref{l2}). The same method may be used for $\bchi_{p}$ by 
differentiating $\bchi_{p}\cdot\bhw = 0$, leading to
\bel{press2}
\frac{D\bchi_{p}}{Dt} = \bchi\times\bchi_{p} + \bq
\hspace{1cm}
\hbox{where}
\hspace{1cm}
\bq = \mu\bchi +  \lambda\,\bchi_{p}
\ee
where $\mu =\mu(\bx,t)$ and $\lambda = \lambda(\bx,t)$ are unknown scalars. Explicitly 
differentiating $\bchi_{p} = \bhw\times P\bhw$ in (\ref{press2}) gives
\bel{alph1}
\bhw\left(\bchi\cdot P\bhw\right) - \alpha_{p}\bchi + \bhw\times\frac{D(P\bhw)}{Dt} = 
\bhw\left(\bchi\cdot P\bhw\right) + \bq\,.
\ee
Using the cross product $\bchi = \bhw\times S\bhw$, this can be manipulated into
\bel{alph2}
\bhw\times\left\{\frac{D(P\bhw)}{Dt} - \alpha_{p}\,S\bhw\right\} = \bq\,,
\ee
which means that
\bel{alph3}
\frac{D(P\bhw)}{Dt} = \alpha_{p}S\bhw + \bq \times\bhw + \varepsilon\bhw
\ee
where $\varepsilon = \varepsilon(\bx,t)$ is a third unknown scalar in addition to $\mu$ 
and $\lambda$ in (\ref{press2}). Thus the Lagrangian derivative of $\alpha_{p} = \bhw\cdot 
P\bhw$ is
\bel{alph4}
\frac{D\alpha_{p}}{Dt} = \alpha\alpha_{p} + \bchi\cdot\bchi_{p} + \varepsilon\,.
\ee
Lagrangian differential relations have now been found for $\bchi_{p}$ and $\alpha_{p}$, but at 
the price of introducing the triplet of unknown coefficients $\mu,~\lambda$, and $\varepsilon$ 
which must adjust in a flow to take the Poisson pressure constraint into account -- they cannot 
be regarded as arbitrary. 
\par\medskip
Dimensional analysis on the various Euler variables governed by equations\footnote{$S$ and $P$ are 
subsumed into the scalars $\alpha,~\alpha_{p},~\chi$ and $\chi_{p}$ so there is no need to consider
them separately.} (\ref{alph4}) for $\alpha_{p}$ and (\ref{press2}) for $\bchi_{p}$ shows that 
$[\,\omega] = T^{-1}$, $[\alpha] = T^{-1}$, $[\chi] = T^{-1}$, whereas $[\alpha_{p}] = T^{-2}$,
$[\chi_{p}] = T^{-2}$.  This means that $[\lambda] = T^{-1}$,~$[\mu] = T^{-2}$ and $[\varepsilon] 
= T^{-3}$. Since the Euler equations possess no other time scale $\mu$, $\lambda$ and 
$\varepsilon$ must be expressible in terms of these units or their ratios
\beq{conj1a}
\mu &=& \mu (\omega,\,\alpha,\,\chi,\,\alpha_{p},\chi_{p})~~~~~\hbox{such~that}~~~~~[\mu] = T^{-2}\,,\\
\lambda &=& \lambda (\omega,\,\alpha,\,\chi,\,\alpha_{p},\chi_{p})
~~~~~\hbox{such~that}~~~~~[\lambda] = T^{-1}\,,\\
\varepsilon &=& \varepsilon (\omega,\,\alpha,\,\chi,\,\alpha_{p},\chi_{p})
\,~~~~~\hbox{such~that}~~~~~[\varepsilon] = T^{-3}\,.
\eeq
Now re-define the triplet such that 
\bel{tetrad1}
\lambda =\alpha + \lambda_{1}\,,\hspace{1cm}
\mu = \alpha_{p} + \mu_{1}\,,\hspace{1cm}
\varepsilon = -2\bchi\cdot\bchi_{p} + \mu_{1}\alpha + \lambda_{1}\alpha_{p}+ \varepsilon_{1}
\ee
where the new triplet is subsumed into the tetrad (the unit tetrad is $\mathbb{I} = [1,\,0]$)
\bel{pidef}
\bPi = \mu_{1}\bzeta +  \lambda_{1}\bzeta_{p} + \varepsilon_{1}\mathbb{I}\,.
\ee
\par\smallskip
\begin{theorem}\label{zetapthm}{\bf[Dynamics of $\bzeta_{p}$]\,:}
The pressure tetrad $\bzeta_{p} = [\alpha_{p},\,\bchi_{p}]$ satisfies
\bel{conj3}
\frac{D\bzeta_{p}}{Dt} = \bzeta\circledast\bzeta_{p} + \bPi\,,
\ee
where the triplet of scalar variables  $\mu_{1},~\lambda_{1}$, and $\varepsilon_{1}$ within 
$\bPi(\bx\,,t)$ is determined by the Poisson equation for the pressure
\bel{Poiss1}
-Tr P = Tr S^{2}-\shalf \omega^{2}\,.
\ee
\end{theorem}
\par\smallskip\noindent
\textbf{Remark:} It has yet to be understood what effect the Poisson pressure constraint 
has on the triplet of scalars $\mu_{1},~\lambda_{1}$, and $\varepsilon_{1}$ within $\bPi$. 
They are not all likely to be zero because, for example, in the simple case of Burgers 
vortex $\alpha = \alpha_{0} = \hbox{const}$\,; $\alpha_{p} = -\alpha_{0}^{2}$\,; $\bchi = 
\bchi_{p}=0~~\Rightarrow~~\varepsilon_{1} = \alpha_{0}^{3}$ and $\lambda_{1} = \mu_{1} = 0$.

\section{\large Ideal MHD}\label{MHD}

As already indicated, these ideas can be pursued for other systems that possess vortex stretching.
The equations of ideal incompressible MHD couple an ideal fluid to a magnetic field $\bB$
\bel{F1}
\frac{D\bu}{Dt} = \bB\cdot\nabla\bB - \nabla p\,,
\ee
\bel{F2}
\frac{D\bB}{Dt} = \bB\cdot\nabla\bu\,,
\ee
together with $\mbox{div}\,\bu = 0$ and $\mbox{div}\,\bB = 0$. The pressure
$p$ in (\ref{F1}) is the combination $p = p_{f} + \frac{1}{2}B^{2}$ where
$p_{f}$ is the fluid pressure. Elsasser variables are defined by combining 
the $\bu$ and $\bB$ fields such that
\bel{F3}
\bv^{\pm} = \bu \pm \bB\,.
\ee
The existence of two velocities $\bv^{\pm}$ means that there are two material 
derivatives (and two time clocks)
\bel{F3a}
\frac{D^{\pm}}{Dt} = \frac{\partial~}{\partial t} + \bv^{\pm}\cdot\nabla\,.
\ee
In terms of these, (\ref{F1}) and (\ref{F2}) can be rewritten as 
\bel{F4}
\frac{D^{\pm}\bv^{\mp}}{Dt} = - \nabla p\,,
\ee
with the magnetic field $\bB$ satisfying
\bel{F5}
\frac{D^{\pm}\bB}{Dt} = \bB\cdot\nabla \bv^{\pm}\equiv \bsig^{\pm}\,,
\ee
together with $\mbox{div}\,\bv^{\pm} = 0$. The $\bsig^{\pm}$-stretching vectors defined in 
(\ref{F5}) obey an Ertel's relation already proved in Gibbon (2002)
\bel{F8}
\frac{D^{\pm}\bsig^{\mp}}{Dt} = - P\bB\,.
\ee
The relations in (\ref{F5}) thus allow us to define
\bel{F9}
\alpha^{\pm} = \bhB\cdot(\bhB\cdot\nabla \bv^{\pm})
\hspace{2cm}
\bchi^{\pm} = \bhB\times(\bhB\cdot\nabla \bv^{\pm})
\ee
having used Moffatt's analogy between the vectors $\bw$ and $\bB$ (Moffatt 1978).
Moreover, because $\bsig^{\pm}$ defined in (\ref{F5}) lie in the plane of the unit 
vectors $\bhB$ and $\bhB\times\bhchi^{\pm}$ we have the decomposition
\bel{stretch1}
\bsig^{\pm} = \alpha^{\pm}\bB + \bchi^{\pm}\times\bB\,.
\ee
Thus it is easy to prove that
\bel{stretch2}
\frac{D^{\pm}|\bB|}{Dt} = \alpha^{\pm}|\bB|\,,
\hspace{2cm}
\frac{D^{\pm}\bhB}{Dt} = \bchi^{\pm}\times\bhB\,,
\ee
which are the equivalent of (\ref{eul4a}) for the Euler equations. The $\alpha^{\pm}$ play 
the role(s) of scalar Elsasser stretching rates, with $\bchi^{\pm}$ as the 
rotation rates. One may also define corresponding variables based upon the Hessian matrix $P$
\bel{F10}
\alpha_{pB} = \bhB\cdot P\bhB\,,
\hspace{2cm}
\bchi_{pB} = \bhB\times P\bhB\,.
\ee
We define the tetrads $\bzeta^{\pm}$ and $\bzeta_{pB}$ as 
follows 
\bel{zetaex2}
\bWB = [0,\,\bB]\,,
\hspace{2cm}
\bzeta^{\pm} = \left[\alpha^{\pm},\,\bchi^{\pm}\right]\,,
\hspace{2cm}
\bzeta_{pB} = \left[\alpha_{pB}\,,\,\bchi_{pB}\right]\,.
\ee
\begin{theorem}\label{thm3}
The magnetic field tetrad $\bWB$ satisfies the two relations
\bel{thm3a}
\frac{D^{\pm}\bWB}{Dt} = \bzeta^{\pm}\circledast\bWB\,,
\ee
\bel{thm3b}
\frac{D^{\mp}~}{Dt}\left(\frac{D^{\pm}\bWB}{Dt}\right) + \bzeta_{pB}\circledast\bWB = 0\,.
\ee
The tetrads $\bzeta^{\pm}$ satisfies the compatibility relation
\bel{thm3c}
\frac{D^{\mp}\bzeta^{\pm}}{Dt} + \bzeta^{\pm}\circledast\bzeta^{\mp} + \bzeta_{pB} = 0\,.
\ee
\end{theorem}
\par\medskip\noindent
\textbf{Proof:} The proof of (\ref{thm3a}) follows immediately from (\ref{F5})
\bel{thm3d}
\frac{D^{\pm}\bWB}{Dt} = \left[0, \frac{D^{\pm}\bB}{Dt}\right] = [0,\bsig^{\pm}]
= \bzeta^{\pm}\circledast\bWB\,,
\ee
where we have used (\ref{stretch1}) at the last step. The proof of (\ref{thm3b}) follows by
combining (\ref{F5}) and (\ref{F8}) together with the fact that $P\bB$ lies in the plane of 
the unit vectors $\bhB$ and $\bhB\times\bhchi_{pB}$. Thus we have the decomposition
\bel{thm3e}
P\bB = \alpha_{pB}\bB + \bchi_{pB}\times\bB\,,
\hspace{1cm}\Rightarrow\hspace{1cm}[0,\,P\bB] = \bzeta_{pB}\circledast\bB\,.
\ee
The proof of (\ref{thm3c}) follows as a compatibility relation between 
(\ref{thm3a}) and (\ref{thm3b}).\hspace{2cm}$\blacksquare$
\par\medskip
Finally MHD-Lagrangian frame dynamics, in the spirit of \S\ref{frame}, needs to be interpreted 
in terms of two sets of ortho-normal vectors $\bhB,\,\bhchi^{\pm},\,(\bhB\times\bhchi^{\pm})$ 
acted on by their opposite Lagrangian time derivatives. After some calculation we find the 
equivalent of (\ref{frameA1})--(\ref{frameA3}) and (\ref{frame2}) is
\beq{frameMHD1}
\frac{D^{\mp}\bhB}{Dt}&=& \bD^{\mp}\times\bhB\,,
\\
\frac{D^{\mp}}{Dt}(\bhB\times\bhchi^{\pm}) &=& \bD^{\mp}\times(\bhB\times\bhchi^{\pm})\,,
\\
\frac{D^{\mp}\bhchi^{\pm}}{Dt} &=& \bD^{\mp}\times\bhchi^{\pm}\,,
\eeq
where the pair of Elsasser Darboux vectors $\bD^{\mp}$ are defined as
\bel{frameMHD2}
\bD^{\mp} = \bchi^{\mp} - \frac{c_{1}^{\mp}}{\chi^{\mp}}\bhB\,,
\hspace{2cm}
c_{1}^{\mp} = \bhB\cdot[\bhchi^{\pm}\times(\bchi_{pB} + \alpha^{\pm}\bchi^{\mp})]\,.
\ee

\vspace{-3mm}
\section{Summary} 

The tetrad reformulation of Euler's equations in this paper appears to be completely natural, 
giving results that are remarkably simple in their expression. It also provides a new hybrid
picture of ideal fluid dynamics in which the Lagrangian fluid parcels carry ortho-normal 
frames, whose rotation velocity depends on the local Eulerian values of the pressure and 
vorticity. These frames are defined by three ortho-normal vectors: (a) along the vorticity; 
(b) along its rate of change following the Lagrangian trajectory; and (c) along the cross 
product of these two unit vectors. This frame is governed by the Darboux vector that has 
components that lie only in the $\bhchi$--$\bhw$ plane.

Remarkably, a picture similar to that for Euler fluids also emerges for magnetic fluids 
described by the ideal MHD equations. The MHD equations have two characteristic velocities, 
corresponding to the two Els\"asser variables. Thus, MHD-Elsasser variables summon two 
Lagrangian characteristics along which the evolutionary equations reduce to ortho-normal 
frame dynamics. Instead of being attached to the vorticity vector, both of these MHD frames 
are attached to the magnetic field vector (Moffatt 1978). The second vector in each moving 
frame is obtained by the rate of change of magnetic field along the Elsasser characteristic. 
The two frames are then completed by taking the cross product of the first two unit vectors 
in each frame. Again the rates of rotation of these Elsasser frames depends on local 
Eulerian properties and the respective Darboux angular velocity vectors have only two 
components in each frame. 
\par\smallskip
An interesting direction of future work would be to numerically monitor the tetrads 
$\bzeta$ and $\bzeta_{p}$ to see how close the relations between them are adhered to. 
To remind the reader of the relation between them, we proved in Theorem 
\ref{thm1} that $\bzeta$ satisfies  
\bel{zintro2a}
\frac{D\bzeta}{Dt} + \bzeta\circledast\bzeta + \bzeta_{p} = 0\,,
\ee
and in Theorem \ref{zetapthm} it was shown that if the triplet of scalars is chosen in a 
certain way then $\bzeta_{p}$ satisfies 
\bel{zintro2b}
\frac{D\bzeta_{p}}{Dt} = \bzeta\circledast\bzeta_{p} + \bPi\,,
\ee
where $\bPi$ is the tetrad, linear in $\bzeta$ and $\bzeta_{p}$, defined in (\ref{pidef}).
Eliminating $\bzeta_{p}$ between (\ref{zintro2a}) and (\ref{zintro2b}) gives
\bel{conj4}
\frac{D^2\bzeta}{Dt^2} + \frac{D\bzeta}{Dt}\circledast\bzeta + \bPi
=  \bzeta\circledast\bzeta\circledast\bzeta\,,
\ee
which is not a completely closed because of the triplet of coefficients 
$\mu_{1},~\lambda_{1},~\varepsilon_{1}$ in $\bPi$ and the need to respect the Poisson 
equation. Through this, the vorticity is related to $P$ and $S$ by
\bel{ev1}
\shalf\omega^{2} = Tr (P + S^{2}) = 
\sum_{i=1}^{3}\left[\lambda_{P}^{(i)} + \left(\lambda_{S}^{(i)}\right)^{2}\right]\,.
\ee
The associated eigenvectors of $P$ and $S$ project onto the the ortho-normal frame 
$(\bhw,\,\bhchi,\,\bhw\times\bhchi)$ to yield the coefficients $\alpha$, $\bchi$, 
$\alpha_{p}$, $c_{1}$ and $c_{2}$. In this ortho-normal basis $P$ has six components 
\bel{ev2}
P = \left(
\begin{array}{lll}
\alpha_{p} & c_{1} & c_{2}\\
c_{1} & \beta & c_{3}\\
c_{2} & c_{3} & \gamma
\end{array}\right)\,,
\ee
but for the present formulation, only $\alpha_{p}$, $c_{1}$ and $c_{2}$ are required from 
$P$ with $\alpha = \bhw\cdot S\bhw$ and $\chi = (\bhw\times\bhchi)\cdot S\bhw$.
\par\smallskip
An alternative way of looking at the coupling between $\bzeta$ and $\bzeta_{p}$ is to 
define the $\pm$-operators as $\bD^{\pm}_{\mathfrak{q}} = D \pm\bzeta\circledast$, where $D = D/Dt$. 
Then (\ref{l1a}), (\ref{zintro2a}) and (\ref{zintro2b}) can be written as
\bel{dpm1}
\bD^{-}_{\mathfrak{q}}\bW  = 0\,,
\hspace{1.5cm}
\bD^{+}_{\mathfrak{q}}\bzeta  = -\bzeta_{p}\,,
\hspace{1.5cm}
\bD^{-}_{\mathfrak{q}}\bzeta_{p}  = -\bPi\,,
\ee
from which we conclude that $\bD^{-}_{\mathfrak{q}}\bD^{+}_{\mathfrak{q}}\bzeta = \bPi$.
\par\smallskip
Finally, recent developments in experimental and numerical capabilities also address the 
hybrid  Lagrangian and Eulerian descriptions of fluid dynamics while theoretical developments 
include the Lagrangian averaged Navier-Stokes-alpha equations. The latter have been reviewed 
in Holm \textit{et al.} (2005). 
The implications for Lagrangian averaging of the hybrid picture of rotating frames in ideal 
fluids presented here will be discussed elsewhere but we note that the rotating frame 
representation may suggest a natural decomposition into fast and slow variables involving 
rapid rotations with slow modulations. 

\par\vspace{2mm}\noindent
\textbf{Acknowledgements:} We thank Marc Brachet of the University of Paris VI, 
Evgenii Kuznetsov of Moscow's Landau Institute, Trevor Stuart of Imperial College, 
Alan O'Neill and Jonathan Matthews of the University of Reading, Mike Cullen of the 
UK Meteorological Office, and Raymond Hide of Oxford University and Imperial College, 
for useful discussions and conversations regarding this problem. The work of 
DDH was partially supported  by the US Department of Energy, Office of Science, 
Applied Mathematical Research. 

\vspace{-4mm}

\bibliographystyle{unsrt}

\end{document}